\documentclass[aps,showpacs,preprint,superscriptaddress]{revtex4}
\usepackage{graphicx}
\usepackage{subfigure}
\usepackage{float}
\usepackage{bm}
\usepackage{txfonts}

\begin{document}

\title{Electron-positron pair creation characterized by the conversion energy}
\author{Ibrahim.Sitiwaldi}
\author{Zi-Liang Li}
\affiliation{Key Laboratory of Beam Technology and Materials Modification of the Ministry of Education, and College of Nuclear Science and Technology, Beijing Normal University, Beijing 100875, China}
\author{Bai-Song Xie \footnote{Corresponding author. Email address: bsxie@bnu.edu.cn}}
\affiliation{Key Laboratory of Beam Technology and Materials Modification of the Ministry of Education, and College of Nuclear Science and Technology, Beijing Normal University, Beijing 100875, China}
\affiliation{Beijing Radiation Center, Beijing 100875, China}

\date{\today}
\begin{abstract}

It is demonstrated that a pair can be characterized under which mechanism it was created according to its conversion energy, a quantity defined as the sum of electron and its conjugate positron's mass-energy, in the study of electron-positron pair creation. The value of this quantity is checked with quantum field theoretical simulations for several field configurations and found that it can describe the creation process with rich physical picture, showing all the creation channels and giving the yields of each channel specifically.  Evenly as a very convenient and powerful detection quantity it can be applicable to some complicated pair creation processes such as that triggered by cooperation of two different photons as well as the dynamically assisted Schwinger mechanism.

\end{abstract}
\pacs{12.20.Ds, 03.65.Pm, 02.60.-x}
\maketitle

\section{Introduction}

In quantum electrodynamics (QED), it was predicted that an external energy associated with a strong field can be convert into matter in the form of electron-positron pairs. Schwinger calculated the rate of pair creation in the constant electric field using a non-perturbative approach \cite{Schwinger} after the earlier works of Sauter \cite{Sauter}, Heisenberg and Euler \cite{Heisenberg}. Although the threshold of field strength is too high to reach up now, it is possible to detect the fascinating phenomena of light to directly convert into matter in the near future with the rapid development of laser technology, for example, the European extreme-light-infrastructure (ELI) program is now advancing \cite{ELI}, therefore, the active theoretical study on electron-positron pair creation is important to support the upcoming experiment.

The Dirac equation for a field-free electron has the positive solutions with energy continuum $E_p\geq m_ec^2$ as well as the negative solutions with energy continuum $E_n\leq -m_ec^2$, we call them positive states and negative states respectively. The QED considers the vacuum in such a condition that all the positive states are empty and all the negative states are occupied. In the presence of external field there are mainly two different mechanisms under which the electron in negative energy continuum can get enough energy from external field to jump into the positive energy continuum and leave behind a "hole" which is also called a positron. The first one is the tunneling effect \cite{Schwinger}, it requires the external field to be the order of $E_c=1.32\times10^{18}\rm{V/m}$ and the second mechanism is photon absorbtion \cite{photon1, photon2, photon3} where the time dependence of external field can induce quantum transitions and the photon energy play a key role.

In past decades many theoretical methods have been employed to study the electron-positron pair creation, such as proper time method \cite{propertime1, propertime2, propertime3}, WKB approximation \cite{WKB} and worldline instanton techniques \cite{world1, world2}. The pair creation processes have been investigated via quantum field theoretical simulations \cite{split, DiracRecent} in the spatial and temporal inhomogeneous external field \cite{CreationDynamics, population, noncompeting, positronic, Tansition, combined, Suotang} as well as via quantum kinetic method \cite{kinetic1, kinetic2} in the spatial homogenous and time depending external field \cite{kinetic0, dynamical2, nuriman1, nuriman2} in recent years.

To understand the underlaying physical picture of pair creation processes is one main purpose of studies in this field. Although the pair creation mechanism was studied by many analytical and numerical approaches, however, there have been lack of a unified method to decide under which mechanism the yields generated, especially when the field configuration is so complicated that any of the mechanisms already predicted could be responsible. Motivated by this, we try to suggest a simple and unified method to reveal clearly the pair creation process by focusing on what mechanism dominates and how the pair creation channels are identified.

In this paper we introduce first the \emph{conversion energy} defined as the sum of mass-energy of electron and its conjugate positron. And then we demonstrate that it is directly associated with the creation mechanism of this pair. To test the feasibility of conversion energy we study the spectrum of conversion energy of created pairs for electron-positron pair creation process in a static and oscillating field via quantum field theoretical simulation. We also simulate the conversion energy for pair creation in a bifrequent and a combined field, to detect the process of cooperation of two different photons as well as dynamically assisted Schwinger mechanism predicted by studies \cite{bif1} and \cite{dynamical0,dynamical1} respectively.

The paper is organized as follows. In Sec.\ref{method} we introduce the conversion energy term and its spectrum after a brief review of the computational framework. In Sec.\ref{result} we give the numerical results and analysis. In the last section we provide the conclusion.

\section{Theoretical Method }\label{method}

\subsection{The quantum field theoretical simulations}

The electron and positron are described by Dirac field operator that satisfies the time-dependent Dirac equation, reads as below in one dimensional space (atomic unit is used in this paper),
\begin{equation}\label{de}
  i\partial\hat{\phi}(z,t)/\partial t=[c\sigma_1\hat{p}_z+\sigma_3 c^{2}+V(z,t)]\hat{\phi}(z,t).
\end{equation}
If we focus on a single spin, the usual four-component spinor wave function can be reduced to only two components and Dirac matrices  reduce to the usual Pauli matrices $\sigma$. The  external force is represented by the scalar potential $ V(z,t) $  that varies only in the $z$ direction, for simplicity we redefine $p_z=p$.

The time-evolved Dirac operator may be expanded using time-independent or time-evolved energy eigenfunctions of the field-free Dirac equation as follows:
\begin{eqnarray}\label{expand}
 \hat{\phi}(z,t)&=&\sum_p \hat b_p(t)W_p(z)+\sum_n \hat d^\dag_n(t)W_n(z) \nonumber\\
  &=&\sum_p \hat b_p(t=0)W_p(z,t)+\sum_n \hat d^\dag_n(t=0)W_n(z,t),
 \end{eqnarray}
where $ W_p(z) $ and $W_n(z) $ represent the field-free energy eigenstates $|p\rangle$ and $|n\rangle$ in the spatial representation at $t=0$,  $ W_p(z,t) $ and $ W_n(z,t) $ satisfy the single-particle time-dependent Dirac equation Eq.(\ref{de}). From Eq.(\ref{expand}) we obtain
 \begin{eqnarray}\label{bb}
 \hat b_p(t)=\sum_{p'}\hat b_{p'}(t=0)\langle p|U(t)|p'\rangle+\sum_{n'}\hat d^\dag_{n'}(t=0)\langle p|U(t)|n'\rangle ,
 \end{eqnarray}
 \begin{eqnarray}\label{dd}
 \hat d^\dag_n(t)=\sum_{p'}\hat b_{p'}(t=0)\langle n|U(t)|p'\rangle+\sum_{n'}\hat d^\dag_{n'}(t=0)\langle n|U(t)|n'\rangle,
 \end{eqnarray}
where the coefficients are the elements of the time-ordered propagator $U(t)=\exp[-i\int^tdt'[c\sigma_1p+\sigma_3c^2+V(z,t')]]$ between energy eigenstates.

The created electrons' spatial density can be obtained from the expectation value of the product of the electronic filed operators,
 \begin{equation}\label{density}
   \rho(z,t)=\langle\langle \rm{vac}||\hat{\phi}_e^\dag(z,t)\hat{\phi}_e(z,t)||\rm{vac}\rangle\rangle
 \end{equation}
 where $\hat{\phi}_e(z,t)=\sum_pb_p(t)W_p(z)$ is the electronic portion of the field operator. Using the commutator relations $[\hat b_p,\hat b_{p'}^\dag]_+=\delta_{p,p'}$ and $[\hat d_n,\hat d_{n'}^\dag]_+=\delta_{n,n'}$ the density can be expressed through the field-free energy eigenstates of the single-particle Hamiltonian as
  \begin{equation}\label{density2}
    \rho(z,t)=\sum_n|\sum_pU_{p,n}(t)W_p(z)|^2,
  \end{equation}
  where $U_{p,n}=\langle p|n(t)\rangle=\langle p|U(t)|n\rangle$ can be computed by using the split operator numerical technique \cite{split, Dirac1}.
   By integrating Eq.(\ref{density2}) over space, we obtain the total number of the created pairs as
  \begin{equation}\label{alnumber}
    N(t)=\int dz\rho(z,t)=\sum_{p,n}|U_{p,n}(t)|^2.
  \end{equation}

\subsection{The conversion energy and its spectrum}

The energy of a created pair can be expected to be equal to the energy it has absorbed from external field during its creation  process of light to convert into matter. Note that it should be a physical process which include the possible inverse process of the matter to convert into energy as long as the energy conservation holds. We take the sum of mass-energy of the electron and its conjugate positron and define it as the \emph{conversion energy},
\begin{equation}\label{conversionenergy}
  E_{p,n}=\sqrt{p^2c^2+c^4}+\sqrt{n^2c^2+c^4},
\end{equation}
where $p$ and $n$ is momentum of the electron and its conjugate positron respectively. Equivalently, one can regard the conversion energy as follows, the electron in the negative state with energy $E_n=-\sqrt{n^2c^2+c^4}$ jump into positive state with energy $E_p=\sqrt{p^2c^2+c^4}$, absorbing energy $E_{p,n}=E_p-E_n$ from external field.

The potential height and photon energy are the characteristic value of energy conversion from light to matter for each pair created by tunneling effect and photon absorbtion respectively, so the conversion energy of a pair is directly associated with its creation mechanism once the strength and oscillating frequency of external field is already known.

The expectation number of a pair of electron with momentum $p$ and positron with momentum $n$ created at time $t$ is $|U_{p,n}(t)|^2$, which is given above. We calculate the conversion energy $E_{p,n}$ and corresponding yields $|U_{p,n}(t)|^2$ for all created pairs and represent them in the form of distribution of pair numbers as a function of the conversion energy, $\rho(E,t)$,
\begin{equation}\label{espectrum}
  \rho(E,t)=\frac{\sum_{E-\Delta E/2}^{E+\Delta E/2}|U_{p,n}(t)|^2}{\Delta E},
\end{equation}
where $\sum_{E-\Delta E/2}^{E+\Delta E/2}$  denotes summation for all $p$ and $n$ satisfy $E-\Delta E/2\leq E_{p,n}\leq E+\Delta E/2$ and $\Delta E$ should be chosen properly, too large or too small $\Delta E$ causes $\rho(E,t)$ nonsensitive or nonlinear. We call $\rho(E,t)$ the \emph{conversion energy spectrum} (CES).

\section{Numerical Result}\label{result}

We simulate pair-creation processes for different field configurations in symmetric potential well and investigate how the CES can reveal the creation mechanism for all pairs specifically.

A symmetric potential well is illustrated as $V(z,t)=V(t)S(z)$ for external field in our simulations, where $V(t)$ is potential height and $S(z)=\{\tanh[(z-D/2)/W]-\tanh[(z+D/2)/W]\}/2$ is the Sauter-type potential well, $W$ is a measure for the width of each edge, $D$ is the total extension of the potential well.

For all simulations in this paper, we take $W=0.3/c$, $D=8/c$, the length of simulation space $L=2.5$ with $N_z=4096$ grid points, total simulation time $t_0=40\pi/c^2$ with $N_t=4000$ grid points.
For the calculation of CES, Eq.(\ref{espectrum}), we take $N_E=1000$ grid points in abscissa $E$ and take $\Delta E=0.04c^2$ to get more sensitive and smooth curves.

\subsection{Pair creation in a static electric field}

 \begin{figure}[H]\suppressfloats
\includegraphics [width=8.5cm]{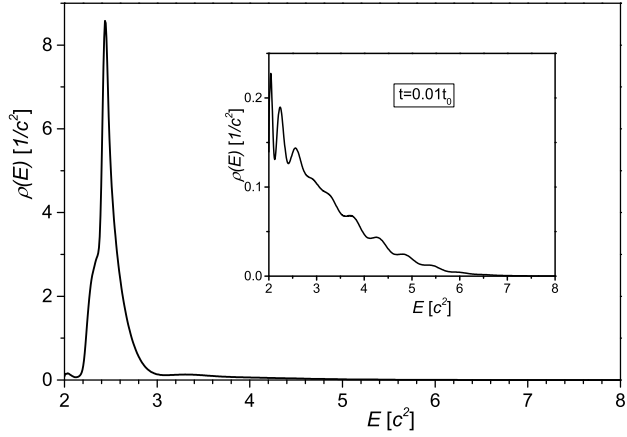}
\caption{\label{static} CES of the final yields in the static field with potential height $V=2.5c^2$. In the inset it is CES shortly after turning on electric field. }
 \end{figure}

In FIG.\ref{static} we display CES of final yields for static field with potential height $V=2.5c^2$, where tunneling effect \cite{Schwinger} is responsible for the creation process. There are a peak corresponding to potential height and most of the pairs are within it. The position of peak is $2.46c^2$ with the agreement of $1.6\%$ with potential height, average conversion energy of total yields is $2.58c^2$ with the agreement of $3.2\%$ with the potential height. We repeat this simulation for the static fields with potential heights $V=3c^2,3.5c^2$ and get graphics very similar to FIG.\ref{static} except the peak positions of $2.94c^2$, $3.44c^2$ respectively. As we mentioned, the potential height is characteristic value of the energy conversion from light to matter, it agrees with the physical picture of tunneling effect very well, the electron in the negative energy continuum  travels through the edge of potential well to join to the positive energy continuum, absorbing energy in the amount of potential height from external field. The conversion energy of a pair created under tunneling process by the static field with potential height $V$ can be expressed as
\begin{equation}\label{height}
E(V)=V.
\end{equation}

The distribution is a bit higher unexpectedly in the left side of the peak, within the area from $2.18c^2$ to $2.38c^2$, it represents that there are some pairs still inside the interaction zone and have lower conversion energy. There are also small amount of distributions out of the peak, these pairs are created in the beginning due to dramatic turning on the external field as we discuss below.

In the inset of FIG.\ref{static} we display CES of created pairs shortly after (when $t=0.01t_0$) turn on the external field. It is very different from the main figure, most of these pairs have too large or too small conversion energy to be created by the tunneling process, indicating there is other creation mechanism. We attribute it to high frequency Fourier components of the external field due to dramatic turning on.

\subsection{Pair creation in a oscillating electric field}

Now we study the pair creation processes for time-depending subcritical electric field,  where the photon absorbtion \cite{photon1,photon2,photon3} can be responsible for the pair creation. The photon energy play a key role and transition amplitude of each order of photon absorbtion can be estimated by the perturbation theory as the appendix of \cite{combined}. We investigate
whether the CES can represent all creation channels triggered by different order of photon absorbtion.

\begin{figure}[H]\suppressfloats
\includegraphics [width=8.5cm]{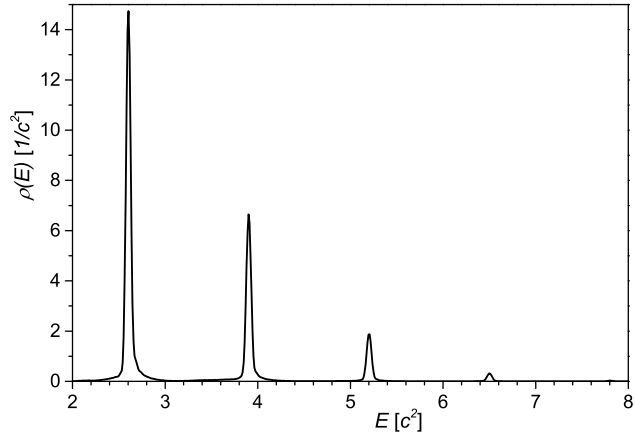}
\caption{\label{13} CES of final yields in the oscillating field with amplitude $V_1=1.47c^2$ and frequency $\omega_1=1.3c^2$.}
\end{figure}

In FIG.\ref{13} we display CES of final yields for the oscillating field  with amplitude $V_1=1.47c^2$ and frequency $\omega_1=1.3c^2$ where the potential height oscillates as $V(t)=V_1\sin(\omega_1t)$. Almost all pairs are distributed within four narrow peaks, the positions of them are $E_2=2.61c^2$, $E_3=3.90c^2$, $E_4=5.21c^2$ and $E_5=6.50c^2$, respectively. These peaks represent the two, three, four and five-photon processes. The conversion energy of a pair created under photon absorbtion by the oscillating field with frequency $\omega_1$ can be expressed as
\begin{equation}\label{photonnumber}
E(\omega_1)=n\omega_1
\end{equation}
where $n$ is the number of absorbed photons and satisfies $E(\omega_1)\geq2c^2$.

The corresponding yields are $60.9\%$, $29.6\%$, $8.19\%$, and $1.32\%$ of total yields $1.73$, respectively where the yields of each process is obtained by integrating corresponding peaks. The yields due to lower order photon absorbtion is more than of higher order photon absorbtion because the chance of absorbing large number of photons is smaller than the chance of absorbing small number of photons in the view of perturbation theory.

It is need to remind that FIG.\ref{13} is similar to multiphoton peaks in \cite{stimulated}, but there are still have significant difference. Instead of momentum of positron we take the exact mass-energy of electron and its conjugate positron to represent how much energy this pair absorbed from external field.

\begin{figure}[H]\suppressfloats
\includegraphics [width=8.5cm]{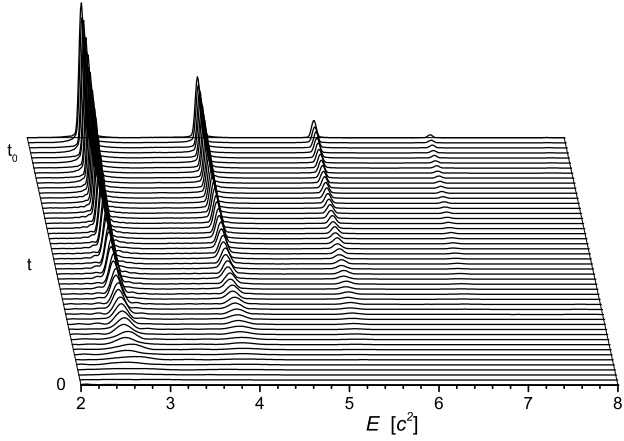}
\caption{\label{waterfall} Waterfall of CES of yields in time $t=(n/50)t_0$  $ (n=1,2,3,...,50)$ for oscillating field with amplitude $V_1=1.47c^2$ and frequency $\omega_1=1.3c^2$. }
 \end{figure}

In order to describe the creation process during all the simulation time we display the waterfall of CES of yields in time $t=(n/50)t_0$ $ (n=1,2,3,...,50)$ for oscillating field with amplitude $V_1=1.47c^2$ and frequency $\omega_1=1.3c^2$ in FIG.\ref{waterfall}. All the peaks are generated from very beginning and the peak positions keep unchanged while the peak heights linearly increase following the simulation time, representing the creation process with constant creation mechanism almost all the interaction time.

\begin{figure}[H]\suppressfloats
\includegraphics [width=8.5cm]{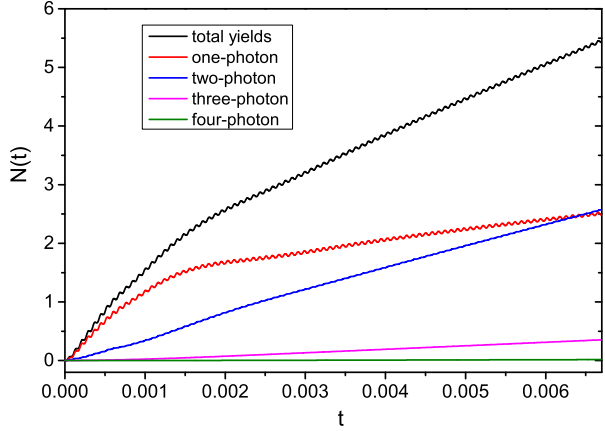}
\caption{\label{FIG4} (Color online) The yields of each creation channel as a function of time in oscillating field with frequency $\omega_1=2.1c^2$ and amplitude $V=1.47c^2$.  }
 \end{figure}

Integrating corresponding peaks in each time step we obtain the yields of each creation channels as a function of time, we find that the temporal behaviour of each creation channel can give some new physical insights. In FIG.\ref{FIG4} we represent time dependence of each creation channel in oscillating field with frequency $\omega_1=2.1c^2$ and amplitude $V=1.47c^2$. The creation rate of total yields decrease during early time and then keeps unchanged. The creation rate of each creation channel in this figure explore that the one-photon process is responsible for the reduction in total creation rate while other channels keep the constant creation rate all the time. The conversion energy of one-photon process is $2.1c^2$, which means the created pairs have very low kinetic energy and little chance to escape from the creation zone, thus particle density in the creation zone and the suppression of Pauli block on corresponding creation process increase until the creation rate so low that the particle number density in creation zone do not increase any more.

It is need to remind that a pair may interact again with the electric field once it is created and absorb more photons to contribute the higher order process so the physical picture of high order process is still not very clear. For example, we are unable to decide that the pairs corresponding to the peak at $E=3.90c^2$ in FIG.\ref{13} absorb three photons during its creation or absorb one more photon after created by the two-photon process.

\subsection{Pair creation in a bifrequent electric field}

The pair creation process triggered by two different photon cooperatively under the bifrequent field with frequency $\omega_1, \omega_2$ was predicted via Boltzmann-Vlasov equation and multiphoton peaks was represented by  \cite{bif1} . We simulate this process to reveal all creation channels triggered by two different photons as well as corresponding yields of each channel.

\begin{figure}[H]\suppressfloats
\includegraphics [width=8.5cm]{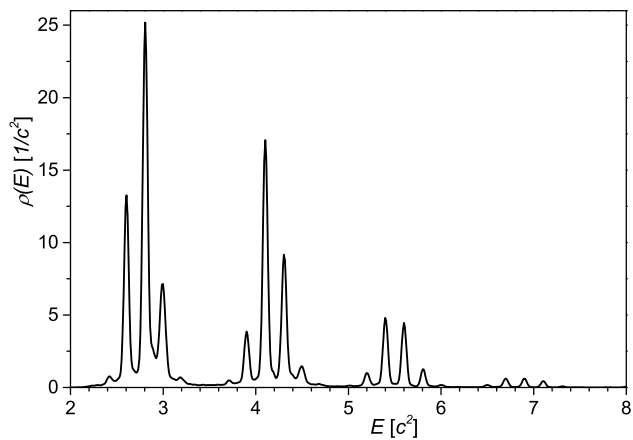}
\caption{\label{1315} CES of final yields in the combination of two oscillating fields with amplitudes $V_1=V_2=1.47c^2$ and frequencies $\omega_1=1.3c^2$, $\omega_2=1.5c^2$. }
 \end{figure}

In FIG.\ref{1315} we display CES of the final yields for bifrequent field composed of two oscillating fields with amplitudes $V_1=V_2=1.47c^2$ and frequencies $\omega_1=1.3c^2$, $\omega_2=1.5c^2$ where the total potential height oscillates as $V_3(t)=V_1\sin(\omega_1t)+V_2\sin(\omega_2t)$. There are many narrow peaks with different heights, indicating the existence of rather complicated creation channels. The two, three, four-photon processes triggered by photon with frequency $\omega_1$ denoted as $\omega_1$, $2\omega_1$, $3\omega_1$ are represented by the peaks at $E=2.61c^2$, $3.90c^2$, $5.20c^2$ respectively, matching with corresponding number of photon energy perfectly. In the same way the two, three, four-photon processes triggered by photon with frequency $\omega_2$ denoted as $\omega_2$, $2\omega_2$, $3\omega_2$ are represented by peaks at $E=3.00c^2, 4.50c^2, 6.01c^2$ respectively.

As we expect we detect the process of photon absorbtion triggered by the cooperation of two different photons here. The corresponding  two-photon process denoted as $\omega_1+\omega_2$, three-photon processes denoted as $2\omega_1+\omega_2$, $\omega_1+2\omega_2$, and four-photon processes denoted as $3\omega_1+\omega_2$, $2\omega_1+2\omega_2$, $\omega_1+3\omega_2$ are represented by the peaks at $E=2.80c^2$, $4.10c^2$, $4.30c^2$, $5.40c^2$, $5.60c^2$, $5.81c^2$ respectively. We neglect to discuss the peaks with rather small heights which are corresponding to higher order processes.

We find some peaks with unexpected positions in FIG\ref{1315}. For example, the creation mechanism of pairs corresponding to the peak at $E=2.42c^2$ can not be regard as a pure photon absorbtion since its corresponding conversion energy can't be expressed as a sum of any positive integer number of $\omega_1$ and $\omega_2$, but matches $3\omega_1-\omega_2$ very well. We explain the creation process as absorbing three photons with frequency $\omega_1$ and emitting one photon with frequency $\omega_2$, but we are still unable to decide that  during creation or after creation it emits a photon. These peaks at $E=3.19c^2$, $3.71c^2$, $4.69c^2$ can also be explained in the similar way with the processes denoted as $3\omega_2-\omega_1$, $4\omega_1-\omega_2$, and $4\omega_2-\omega_1$ respectively. The underlying physical mechanism of these findings need to study further in future.

The conversion energy of a pair created in the bifrequent field with frequency $\omega_1, \omega_2$  can be expressed as
\begin{equation}\label{allmomentum}
E(\omega_1,\omega_2)=n_1\omega_1+n_2\omega_2
\end{equation}
where $n_1,n_2$ are integer satisfying $E(\omega_1,\omega_2)\geq2c^2$.

The final yields is $7.04$, far more than the sum of the final yields due to each field with single frequency individually, which are $1.73$ and $1.77$ respectively, according to our simulations. The creation process is represented with all existing creation channels by CES so we can explain this enhancement easily. Besides the individual creation channels due to each single oscillating field, the extra creation channels are available which are triggered by two different photons cooperatively, we believe that these extra channels are responsible for enhancement in the creation rate due to combination of two oscillating fields.

\subsection{Pair creation in the combined electric field}

We simulate pair creation process for combined field in terms of conversion energy to detect the dynamically assisted Schwinger mechanism which was predicted by study \cite{dynamical0} and to describe it with more details.

\begin{figure}[H]\suppressfloats
\includegraphics [width=8.5cm]{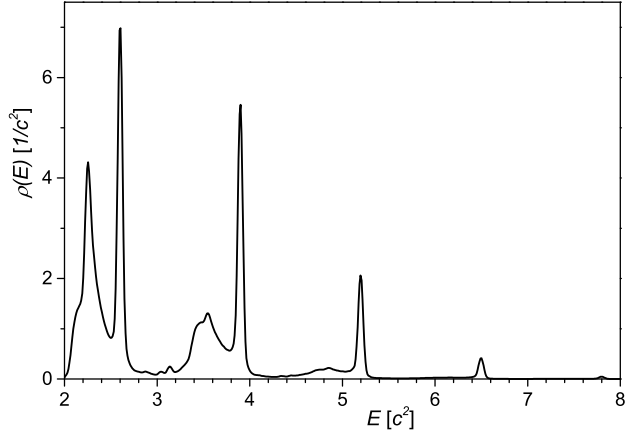}
\caption{\label{combined} CES of final yields in the combination of static field with potential height $V_0=1c^2$ and oscillating field with amplitude $V_1=1.47c^2$, frequency $\omega_1=1.3c^2$. }
 \end{figure}

In FIG.\ref{combined} we display CES of final yields in the combination of static field with potential height $V_0=1c^2$ and oscillating field with amplitude $V_1=1.47c^2$, frequency $\omega_1=1.3c^2$. There are four narrow peaks at $E=2.61c^2$, $3.90c^2$, $5.20c^2$,  $6.50c^2$ representing two, three, four, and five-photon processes denoted as $2\omega_1$, $3\omega_1$, $4\omega_1$, $5\omega_1$. As we expect, we detect the processes of dynamically assisted Schwinger mechanism denoted as $\omega_1+V_0$, $2\omega_1+V_0$, $3\omega_1+V_0$, which are represented by the peaks at $E=2.25c^2$, $3.55c^2$, $4.86c^2$ respectively. The conversion energy of a pair created in combination of static field with potential height $V_0$ and oscillating field with frequency $\omega_1$ can be expressed as
\begin{equation}\label{photonheight}
E(\omega_1,V_0)=n\omega_1+kV_0
\end{equation}
where $k=1$ for dynamical assisted Schwinger process and $k=0$ for pure photon absorbtion, $n$ is integer satisfying $E(\omega_1,V_0)\geq2c^2$.

The total yields is $2.63$, the creation channels associated with tunneling assisted photon absorbtion are responsible for the enhancement compared to $1.73$, the yields of oscillating field individually.

\section{conclusion and discussion}

In the frame of computational simulation we reexamine the problem of electron-positron pair creation in different external background fields. We introduce a simple but useful physical quantity, conversion energy, as the sum of the electron and its conjugate positron's mass-energy to reveal the creation mechanism for all created pairs as well as to detect the various pair creation channels and the corresponding yields.

The conversion energy can not only reveal the pair creation characteristics in the cases of pure tunneling mechanism in a static and pure multiphoton mechanism in an oscillating field  but also unveil the complicated features in the cases of mixed mechanism triggered by the cooperation of two different photons in fields with two different frequencies and the dynamically assisted Schwinger mechanism in the combined field with one static and the other oscillating.

A pair can be identified how it was created by its conversion energy represented as Eq.(\ref{height}), Eq.(\ref{photonnumber}), Eq.(\ref{allmomentum}), and Eq.(\ref{photonheight}), in the computational simulations. By this way it can unveil the pair creation process with more physical insight. Especially it is more powerful to distinguish and identify the certain creation channels individually for the complicated fields configuration.  It is also believed that the concept of conversion energy may be applicable to the other kind of pair creation process, for example, in the laser-nucleon collision as well in the laser-electron beam collision.

\begin{acknowledgments}
We enjoyed several helpful discussions with Suo Tang, and Feng Wan. This work was supported by the National Natural Science Foundation of China (NSFC) under Grant No. 11475026.

\end{acknowledgments}


\begin{thebibliography}{99}\suppressfloats

\bibitem{Schwinger}
J. S. Schwinger, Phys. Rev. 82 (1951) 664.

\bibitem{Sauter}
F. Sauter, Z. Phys. 69 (1931) 742.

\bibitem{Heisenberg}
W. Heisenberg, H. Euler, Z. Phys. 98 (1936) 714.

\bibitem{ELI}
http://www.eli-beams.eu/.

\bibitem{photon1}
E. Brezin and C. Itzykson, Phys. Rev. D 2 (1970) 1191.

\bibitem{photon2}
V. S. Popov, JETP Lett. 13 (1971) 185.

\bibitem{photon3}
R. Alkofer, M. B. Hecht, C. D. Roberts, S. M. Schmidt, and D. V. Vinnik, Phys. Rev. Lett. 87 (2001) 193902.

\bibitem{propertime1}
R. G. Newton, Phys. Rev. 96 (1954) 523.

\bibitem{propertime2}
W. Y. Tsai and A. Yildiz, Phys. Rev. D 8 (1973) 3446.

\bibitem{propertime3}
V. N. Baier, V. M. Katkov, and V. M. Strakhovenko, Sov. Phys. JETP 40 (1974) 225.

\bibitem{WKB}
S. P. Kim and D. N. Page, Phys. Rev. D 65 (2002) 105002.

\bibitem{world1}
G. V. Dunne and C. Schubert, Phys. Rev. D 72 (2005) 105004.

\bibitem{world2}
G. V. Dunne, Q. H. Wang, H. Gies, and C. Schubert, Phys. Rev. D 73 (2006) 065028.



\bibitem{split}
Andr\'{e} D. Bandrauk, and Hai Shen, J. Chem. Phys. 99 (1993) 1185.

\bibitem{DiracRecent}
T. Cheng, Q. Su, and R. Grobe, Cont. Phys. 51 (2010) 315.

\bibitem{CreationDynamics}
P. Krekora, K. Cooley, Q. Su, and R. Grobe, Phys. Rev. Lett. 95 (2005) 070403.

\bibitem{population}
Y. Liu, M. Jiang, Q. Z. Lv, Y. T. Li, R. Grobe, and Q. Su, Phys. Rev. A 89 (2014) 012127.

\bibitem{noncompeting}
Q. Z. Lv, Y. Liu, Y. J. Li, R. Grobe, and Q. Su, Phys. Rev. Lett. 111 (2013) 183204.

\bibitem{positronic}
Y. Liu, Q. Z. Lv, Y. T. Li, R. Grobe,and Q. Su, Phys. Rev. A 91 (2015) 052123.

\bibitem{Tansition}
M. Jiang, Q. Z. Lv, Z. M. Sheng, R. Grobe, and Q. Su, Phys. Rev. A 87 (2013) 042503.

\bibitem{combined}
M. Jiang, W. Su, Z. Q. Lv, X. Lu, Y. J. Li, R.Grobe, and Q. Su, Phys. Rev. A 85 (2012) 033408.

\bibitem{Suotang}
Suo Tang, Bai-Song Xie, Ding Lu, Hong-Yu Wang, Li-Bin Fu, and Jie Liu, Phys. Rev. A 88 (2013) 012106.


\bibitem{dynamical0}
R. Sch\"{u}tzhold, H. Gies, and G. Dunne, Phys. Rev. Lett. 101 (2008) 130404.

\bibitem{dynamical1}
G. V. Dunne, H.Gies, R. Sch\"{u}tzhold, Phys. Rev. D 80 (2009) 111301.

\bibitem{bif1}
I. Akal, S. Villalba-Ch\'{a}vez, and C. M\"{u}ller, Phys. Rev. D 90 (2014) 113004.



\bibitem{kinetic1}
R. Alkofer, et al., Phys. Rev. Lett. 87 (2001) 193902.

\bibitem{kinetic2}
C. D. Roberts, S. M. Schmidt, D. V. Vinnik, Phys. Rev. Lett. 89 (2002) 153901.



\bibitem{kinetic0}
F. Hebenstreit, R. Alkofer, G. V. Dunne, H. Gies, Phys. Rev. Lett. 102 (2009) 150404.


\bibitem{dynamical2}
M. Orthaber, F. Hebenstreit, R. Alkofer, Phys. Lett. B 698 (2011) 80.


\bibitem{nuriman1}
A. Nuriman, B. S. Xie, Z. L. Li, and D. Sayipjamal, Phys. Lett. B 717 (2012) 465.

\bibitem{nuriman2}
A. Nuriman, Z. L. Li, and B. S. Xie, Phys. Lett. B 726 (2013) 820.

\bibitem{stimulated}
T. Heinzl, A. Ilderton, M. Marklund, Phys. Lett. B 692 (2010) 250.

\end{thebibliography}
\end{document}